\documentclass{article}
\usepackage{spconf,amsmath,graphicx}
\usepackage{multirow}
\usepackage{setspace}
\usepackage{tabularx, booktabs, makecell}
\usepackage{hyperref}

\usepackage{cite}
\usepackage{verbatim}

\title{Style modeling for Multi-Speaker Articulation-to-Speech}
\name{Miseul Kim \qquad Zhenyu Piao \qquad Jihyun Lee \qquad Hong-Goo Kang}
\address{Department of Electrical and Electronic Engineering, Yonsei University, Seoul, South Korea}

\begin{document}
%
\maketitle

\begin{abstract}

In this paper, we propose a neural articulation-to-speech (ATS) framework that synthesizes high-quality speech from articulatory signal in a multi-speaker situation. 
Most conventional ATS approaches only focus on modeling contextual information of speech from a single speaker's articulatory features.
To explicitly represent each speaker's speaking style as well as the contextual information, our proposed model estimates style embeddings, guided from the essential speech style attributes such as pitch and energy.
We adopt convolutional layers and transformer-based attention layers for our model to fully utilize both local and global information of articulatory signals, measured by electromagnetic articulography (EMA). 
Our model significantly improves the quality of synthesized speech compared to the baseline in terms of objective and subjective measurements in the Haskins dataset.

  
\end{abstract}

\begin{keywords}
Style-modeling, Multi-speaker, Speech synthesis, Articulation-to-speech
\end{keywords}

\vspace{-4pt}
\section{Introduction}
\label{sec:intro}


An articulation-to-speech (ATS) system aims to generate intelligible human speech from articulatory movements. 
Recently, these systems have gained an emerging popularity as an alternative communication tool for people having difficulties in producing the intelligible speech~\cite{chen2021ema2s} such as laryngectomees~\cite{gates1982current}.
As one of the essential parts of silent speech interface (SSI)~\cite{denby2010silent}, ATS system can also bridge the connection between brain and speech signal~\cite{anumanchipalli2019speech}.

Articulatory movements are represented as digital signals which are recorded from places of articulation such as lips, tongue, and palate.
Since these kinematic recordings contain rich phonetic information of speech~\cite{kauffman1976articulation, volaitis1992phonetic}, there have been several statistical approaches that synthesizing human voice from articulatory signals.
Gaussian mixture model (GMM)-based ATS synthesizers are proposed in~\cite{toda2004mapping, aryal2013articulatory} to estimate Mel-Frequency Cepstral Coefficients (MFCCs) from articulations. 
In \cite{picart2011continuous}, the effects of two different articulation styles are investigated from synthetic acoustic feature from Hidden Markov Model (HMM). 

Thanks to recent advances in deep learning technologies, the performance of ATS systems has significantly improved compared to the conventional signal processing-based ones~\cite{gaddy-klein-2021-improved, kong2020hifi, cao2022speaker, cao2018articulation, taguchi2018articulatory}. Nevertheless, there are some limitations in those approaches.
First, Most of the previous works synthesize speech waveforms by solely utilizing the contextual information in articulatory signals~\cite{chen2021ema2s, csapo2020ultrasound}. 
Thus, there is still room for improvement of the sound fidelity
by providing a speaker's representative acoustic features (e.g., pitch and energy) to the
synthesis process.
Second, Most of the previous ATS systems focus on
a single-speaker synthesis scenario. 
Hence it could be extended to a multi-speaker scenario with a unified system.

In this work, we propose a novel ATS system that estimates a target speaker's speaking style as well as contextual information from a given EMA signal.
As we explicitly model speaking styles of different speakers, our system can support the multi-speaker voice generation scenario with a sole network.
The entire network architecture is composed of four modules: \textit{style encoder, mel generator, and pitch and energy predictor}.
They are built on convolutional neural networks (CNNs) and transformer-based attention layers to effectively capture local and global characteristics from articulatory representations.
Specifically, the synthesis flow of the proposed model is as follows.
First, a style embedding is estimated from articulatory features with one-hot speaker embedding, using a style encoder referenced by 
pitch and energy.
The mel generator then synthesizes mel-spectrum from EMA signal with the style embedding.
Finally, a neural vocoder generates raw waveform from the mel-spectrum.
By doing so, we can synthesize a highly intelligible speech with speaker-specific characteristics from EMA signal under the multi-speaker scenario. Audio samples are available online\footnote{https://miseulkim.github.io}.

Our paper is organized as follows. 
In section \ref{sec:Related works}, we summarize some previous approaches related to our proposed model. 
We describe the detailed architecture of our model in section \ref{sec:proposed_model}. 
Experimental setup and results are presented in section \ref{sec:experiments} and \ref{sec:experimental_results}.
Finally, we conclude the paper in section \ref{sec:conclusion}.

\vspace{-8pt}
\section{Related work}
\label{sec:Related works}

\subsection{Neural articulation-to-speech (ATS) synthesis}
\label{ssec:ats}


Blessed by the development of deep neural networks, neural ATS systems have significantly advanced performances.
Those systems commonly estimate acoustic features such as mel-spectra~\cite{chen2021ema2s, csapo2020ultrasound}, MFCCs~\cite{gaddy-klein-2021-improved}, and vocoder parameters~\cite{cao2018articulation, taguchi2018articulatory}
and use them as intermediate features for waveform synthesis.
To construct high-quality neural ATS systems, they encode high-dimensional latent representations from input articulatory features using neural architectures such as CNNs~\cite{csapo2020ultrasound} or recurrent neural networks (RNNs)~\cite{chen2021ema2s, cao2018articulation, taguchi2018articulatory}.

Introducing neural vocoders is also one of strategies to enhance the quality of synthesized waveform~\cite{chen2021ema2s, gaddy-klein-2021-improved, csapo2020ultrasound}.
These approaches achieve modeling of the contextual information inherent in EMA signal, but there is still room for more human-like speech.
For this problem, \cite{katsurada20_interspeech} utilizes d-vector to represent speaker information.
However, the extracted speaker representation lacks time-varying speaker characteristics because a time-averaging pooling is applied to calculate the d-vector.

In our work, we also adopt the advanced neural architectures: CNNs, Transformer, and neural vocoder to synthesize high-fidelity speech.
Also, to generate more natural speech, we estimate a speaker's time-varying speaking style from articulatory features and utilize it for the synthesis process.


\subsection{Style modeling in text-to-speech (TTS)} 
\label{ssec:tts}
Modeling of a speaker's speaking style has been widely studied in TTS systems.
Fastspeech2~\cite{ren2020fastspeech} utilizes speaker-specific prosody features such as pitch and energy to improve the naturalness of synthesized speech. 
To explicitly model a speaker's speaking style as an embedding vector independently of text, \cite{wang2018style} proposes global style tokens (GSTs) and verifies their effectiveness on various style-changing tasks including multi-speaker TTS. 
\cite{min2021meta} uses a mel-style encoder which is trained to capture speaker identity and prosody and achieves a high-quality multi-speaker adaptive TTS system.

Similar to previous work, we explicitly model a speaker's representative style and apply our ATS system to the multi-speaker scenario.
The difference from aforementioned approaches is that we extract style information only from articulatory features.
We use a style encoder motivated from \cite{min2021meta}, which is trained referencing additional acoustic features such as pitch and energy as in \cite{ren2020fastspeech}.

\section{Proposed Model}
\label{sec:proposed_model}
\vspace{-8pt}

\begin{figure}[t]
\centering
\centerline
{\includegraphics[width=8.3cm]{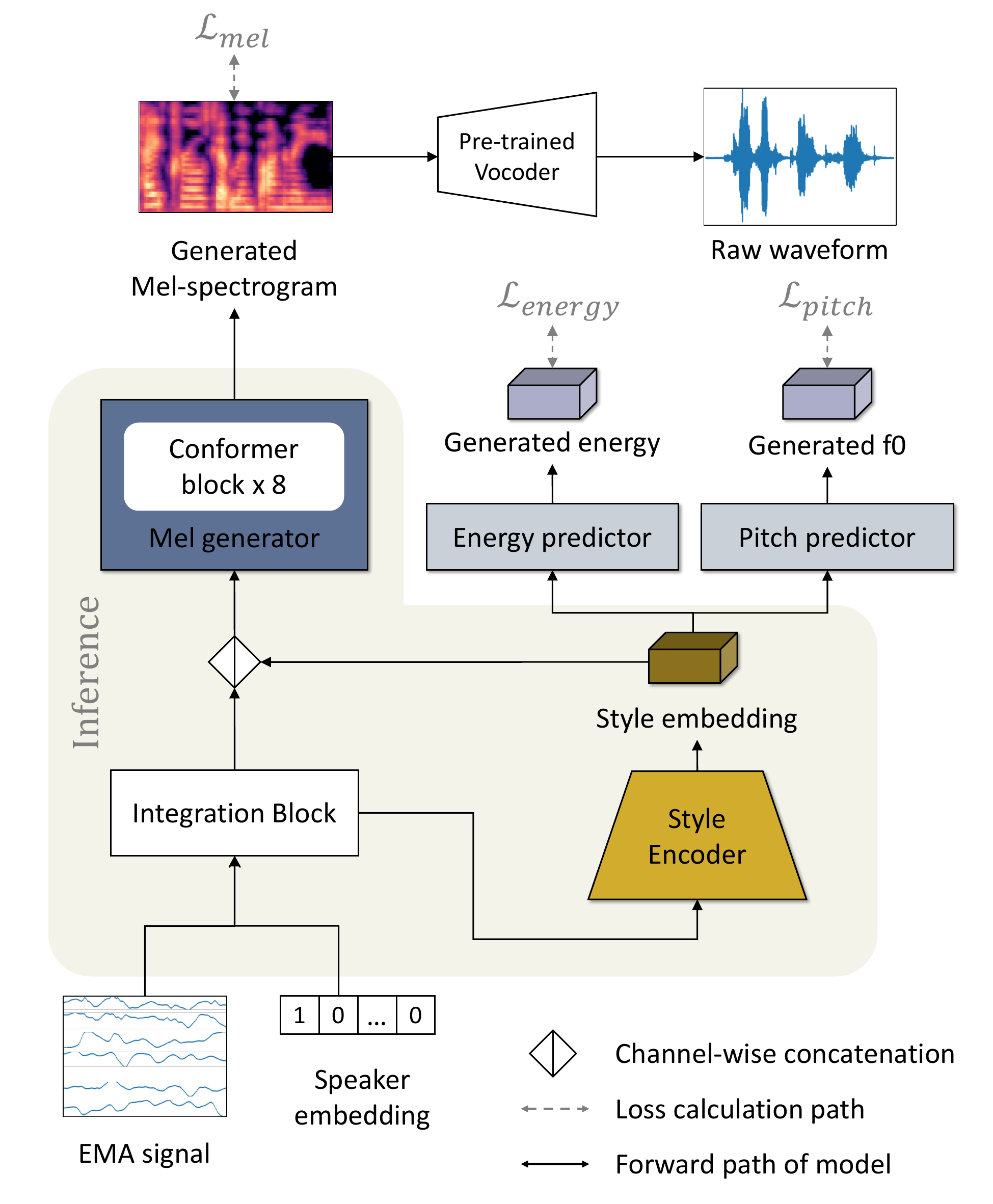}}
\caption{Overall framework of the proposed model. For inference step, only the part inside the light-yellow colored area is utilized.}   
\label{fig:proposed model}
\end{figure}

In this section, we describe the architecture of the proposed model in detail. As shown in Figure~\ref{fig:proposed model}, our model consists of four modules: mel generator, style encoder, pitch and energy predictor. The computation flow is given as follows.

First, to provide a target speaker's information to perform a multi-speaker ATS scenario, one-hot speaker embedding concatenated with EMA signal in channel-wise is given as an input to the integration block. Next, this combined vector passes through a CNN and gated-linear unit (GRU) layer with residual connection (denoted as integration block). 
Then, the 256-dimensional output vector of the integration block is fed into the style encoder to create a style embedding.
Continually, the obtained style embedding is used as an input of the pitch and energy predictor in the training process and then pitch and energy are extracted from. 
The style embedding is also concatenated with the output embedding of the integration block to provide style information for the mel generator to generate mel-spectrum. Finally, the generated mel-spectrogram passes through the pre-trained neural vocoder to synthesize speech waveform.

\begin{table*}[t]
\centering
\caption{Comparison of baseline and proposed model in MOS with confidence intervals of 95\%. Higher score indicated better quality}
\vspace{3pt}
\label{tab:subjective}
\resizebox{\linewidth}{!}
{
\begin{tabular}{c|cccc|cccc|c}
\Xhline{2\arrayrulewidth}
\multirow{2}{*}{} & \multicolumn{4}{c|}{Male speaker}                                                                                                                    & \multicolumn{4}{c|}{Female speaker}                                                                                                                      & \multirow{2}{*}{Average} \\ \cline{2-9}
                  & \multicolumn{1}{c}{M01}                & \multicolumn{1}{c}{M02}                & \multicolumn{1}{c}{M03}                & M04                & \multicolumn{1}{c}{F01}                & \multicolumn{1}{c}{F02}                & \multicolumn{1}{c}{F03}                & F04                &                        \\ \hline
Baseline          & \multicolumn{1}{c}{2.82±0.11} & \multicolumn{1}{c}{2.21±0.11} & \multicolumn{1}{c}{2.81±0.12} & 2.75±0.12 & \multicolumn{1}{c}{2.70±0.11} & \multicolumn{1}{c}{2.58±0.11} & \multicolumn{1}{c}{2.75±0.11} & 2.71±0.11 & 2.67±0.04     \\ \hline
Proposed          & \multicolumn{1}{c}{\textbf{3.25±0.12}}               & \multicolumn{1}{c}{\textbf{2.63±0.11}}               & \multicolumn{1}{c}{\textbf{3.27±0.12}}               & \textbf{3.33±0.11}              & \multicolumn{1}{c}{\textbf{3.35±0.11}}               & \multicolumn{1}{c}{\textbf{3.12±0.12}}               & \multicolumn{1}{c}{\textbf{3.62±0.11}}               & \textbf{3.36±0.11}               &\textbf{3.24±0.04}
              \\ \hline
\end{tabular}}
\vspace{-9pt}
\end{table*}

\subsection{Style encoder}
EMA signals only contain a limited amount of information about the acoustic properties of supralaryngeal configuratons (e.g., formants).
However, human voice is composed of various elements such as formant, pitch, energy and prosody information depending on different characteristics of each speaker. Therefore, we adopt a style encoder to obtain style vector which possesses an additional acoustic information to generate expressive mel-spectrums from the input EMA signals. 

Based on the assumption that EMA is correlated with both contextual and speaking style-related information, we directly estimates speech style embedding from EMA by referencing pitch and energy component. Similar to the structure of mel style encoder in \cite{min2021meta}, our style encoder consists of one linear, two CNN blocks with residual connection, one multi-head attention and one linear layer sequentially. 

\subsection{Pitch and Energy predictor}
For pitch and energy predictor, we utilize the architecture of variance predictor refers to \cite{ren2020fastspeech}, which consists of two CNN layers and one linear layer. Each predictor generates $1*T$ frame-level pitch and energy output vector respectively. 
To minimize the distance between the output of each module and ground-truth target, L1 loss function is utilized for training. Also, to reduce the training error, target pitch and energy are represented as a decibel (dB) unit. 

\subsection{Mel generator}
The mel generator is designed to generate high-quality mel-spectrum from a given articulatory movement and a style embedding. For modeling both the content-based global information and local features effectively from input signal, we borrows the backbone of conformer \cite{gulati2020conformer}, which integrates CNN and transformer layer. For more details, our conformer block consists of one convolution block with residual connection, transformer self-attention layer with 4 head and two linear layers in order. Total 8 conformer blocks are stacked for constructing generator.

In conclusion, total training loss function $L_{total}$ is defined as below with weight parameters $\lambda_{mel}=0.8, \lambda_{pitch}=0.1$ and $\lambda_{energy}=0.1$ respectively.
\begin{equation}
    L_{total}=\lambda_{mel}L_{mel}+\lambda_{pitch}L_{pitch}+\lambda_{energy}L_{energy}.
\end{equation}
As depicted in yellow-colored territory in Figure \ref{fig:proposed model}, only the integration block, style encoder and mel-spectrum generator are used to generate mel-spectrogram at the inference stage.

\section{Experimental setup}
\label{sec:experiments}

\subsection{Dataset}
\label{sec:dataset}
Haskins
public dataset is utilized for our ATS task. The data consists of correspondingly recorded EMA and audio file for 8 native American English speakers. Average recording duration for per speaker is about 55 minutes. Kinematic data and audio samples are recorded in 100 Hz and 44.1 kHz sampling rate respectively.
Among the recorded 8 point movements, 6 electrodes part are used for our task, TR (tongue rear), TB (tongue blade), TT (tongue tip), UL (upper lip), LL (lower lip), jaw (lower medial incisors). Each sensor position includes trajectories orientation, X (posterior to anterior), Y (right to left), Z (inferior to superior). Therefore, EMA vector of model's input consists of 18-dimensions (6 electrode * 3 dimension).

\begin{table}[t!]
\caption{STOI, PESQ, MCD and CER for each model. Summation symbol '+' means that embedding components are added on the upper row system.}
\vspace{4pt}
\begin{spacing}{1.1}
\resizebox{\linewidth}{!}
{%
\centering
{\scriptsize
\begin{tabular}{c|cccc}
\Xhline{2\arrayrulewidth}
Model               & STOI ↑& PESQ ↑& MCD ↓& CER(\%) ↓ \\ \hline
Baseline            & 0.663 & 1.154 & 6.809 & 31.14    \\ \hline
Mel generator      & 0.689 & 1.181 & 6.513 & 26.62     \\
+ Speaker embedding & 0.693 & 1.186 & 6.474 & \textbf{25.68}    \\
++ Style embedding  & \textbf{0.697} & \textbf{1.193} & \textbf{6.436} & 26.01     \\ \hline
\end{tabular}

}}
\vspace{-8pt}

\label{tab:objective}
\end{spacing}
\end{table}

\vspace{-4pt}
\subsection{Implementation details}
\label{sec:implementation_details}

\textbf{Pre-processing} Because kinematic data and audio samples are recorded in substantially different sampling rates, we apply interpolation to EMA samples for matching the number of sample as the one with target acoustic feature. 40-dimentional mel-spectrogram is the target for our approaches, which is extracted with 1024 sample window length overlapping by 25\%, considering the low sampling rate of EMA. 
And the target audio is contaminated by background noise sound, hence we remove the noise using speech enhancement module \cite{sainburg2020finding}. Also, we trim silence part from both of EMA and audio samples to prevent the model confusion during training.

\noindent\textbf{Baseline} To perform a comparsion with SOTA transformer-based model, we build a baseline~\cite{gaddy-klein-2021-improved}, which stacks three residual convolution blocks and six Transformer layers in order. Different from \cite{gaddy-klein-2021-improved} which estimates MFCC features, we train the baseline to estimate mel-spectrum from the articulatory features. For training criteria, we adopt L1 loss to calculate the distance of mel-spectrum between generated and ground truth.

\noindent\textbf{Vocoder setup} For obtaining the raw waveform from mel-spectrogram, hifi-GAN~\cite{kong2020hifi} neural vocoder is adopted.  
Since the number of audio sample data in haskin is insufficient for training, we first pre-train the hifi-GAN with LJspeech dataset~\cite{ljspeech17} and then fine-tune the model with haskins.

\begin{figure}[t]
\centering
\centerline{\includegraphics[width=8.4cm]{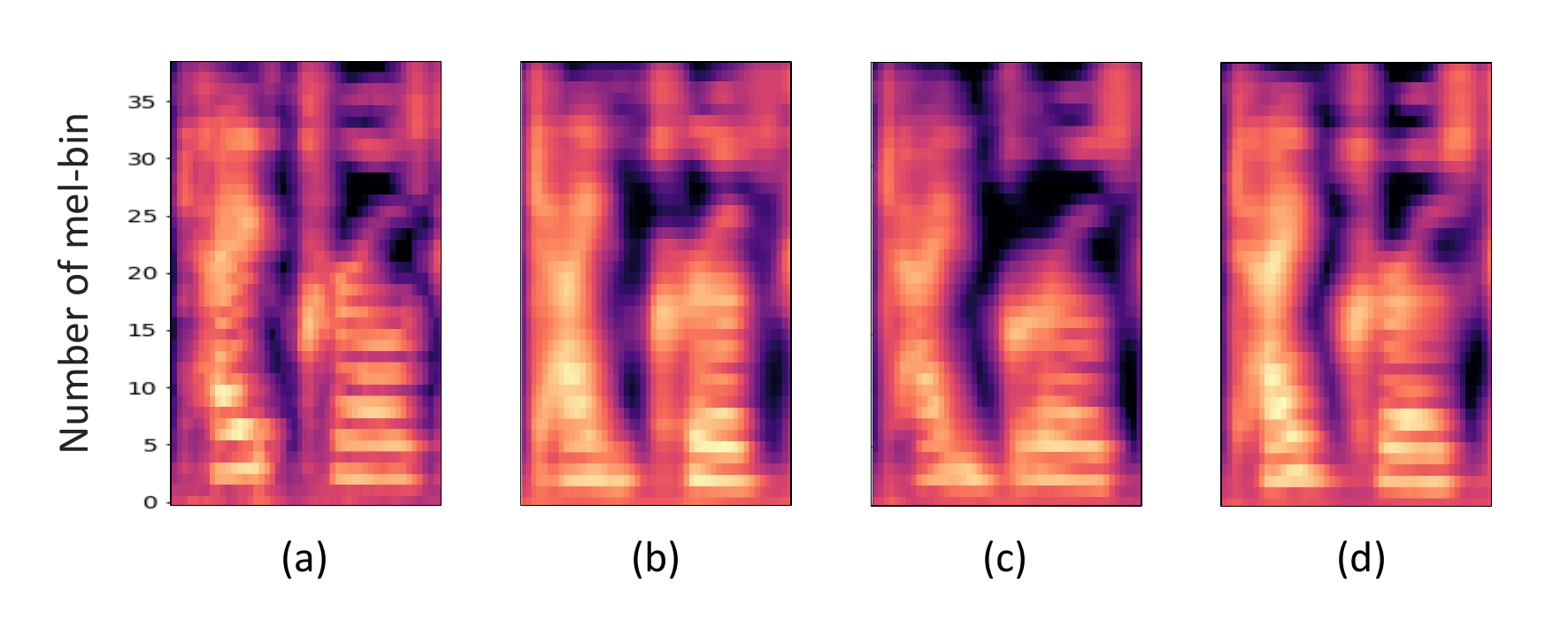}}
\caption{Mel spectrum of reference and estimated samples for three systems. (a): Target spectrum (b): Baseline (c): Mel generator with speaker embedding (d): Mel generator with speaker embedding and style embedding.}   
\vspace{-8pt}
\label{fig:analysis}
\end{figure}
\vspace{-16pt}

\subsection{Evaluation metric}
\label{sec:evaluation_metric}
We evaluate our results in terms of subjective and objective metrics. Specifically, for subjective quality evaluation, we conduct MOS (mean opinion
score) test. 
For objective quality evaluation, we utilize four criteria, mel-cepstral distortion (MCD), perceptual evaluation of speech quality (PESQ), short-time objective intelligibility (STOI) and character error rate (CER). We use a pre-trained ASR model--- Wav2Vec~\cite{baevski2020wav2vec} to compute CER.
\vspace{-8pt}

\section{Experimental results}
\label{sec:experimental_results}

\subsection{Subjective quality measurements}
To demonstrate the fidelity of the synthetic speech, we conduct a mean opinion score (MOS) test between the baseline and proposed model. Total 20 respondents are asked to give discrete scores from 1 to 5. We randomly select 20 sentences from the test dataset per speaker and system. 

As shown in the Table \ref{tab:subjective}, our proposed network achieves much superb score than the baseline for all speakers cross different genders. 
The experimental results fully demonstrate the efficiency of proposed model of generating more high-fidelity speech from articulatory signals than the baseline.
\vspace{-8pt}

\subsection{Objective quality measurements}
We summarize objective quality of synthetic speech in Table~\ref{tab:objective}.
Our conformer-based architectures show much superb performance than baseline model in terms of MCD, PESQ, STOI. 
Experimental result shows that guiding the model with one-hot speaker embedding could further improve the quality of the synthetic speech. 
We also observe that adding both speaker and style embedding to the network achieves the best performance among the networks. Thus, we conclude that the two embeddings are definitely effective for enabling the network to generate high-quality speech. 
However, we find that CER performance slightly decline when a style embedding given to the network.
It may result from the synthetic speech conditioned on the style embedding has more prosody variation compared to the baseline, therefore, it could suffer from the distortion in phoneme characteristics~\cite{rani2012error}.

\begin{figure}[t]
\centering
\centerline{\includegraphics[width=8.5cm]{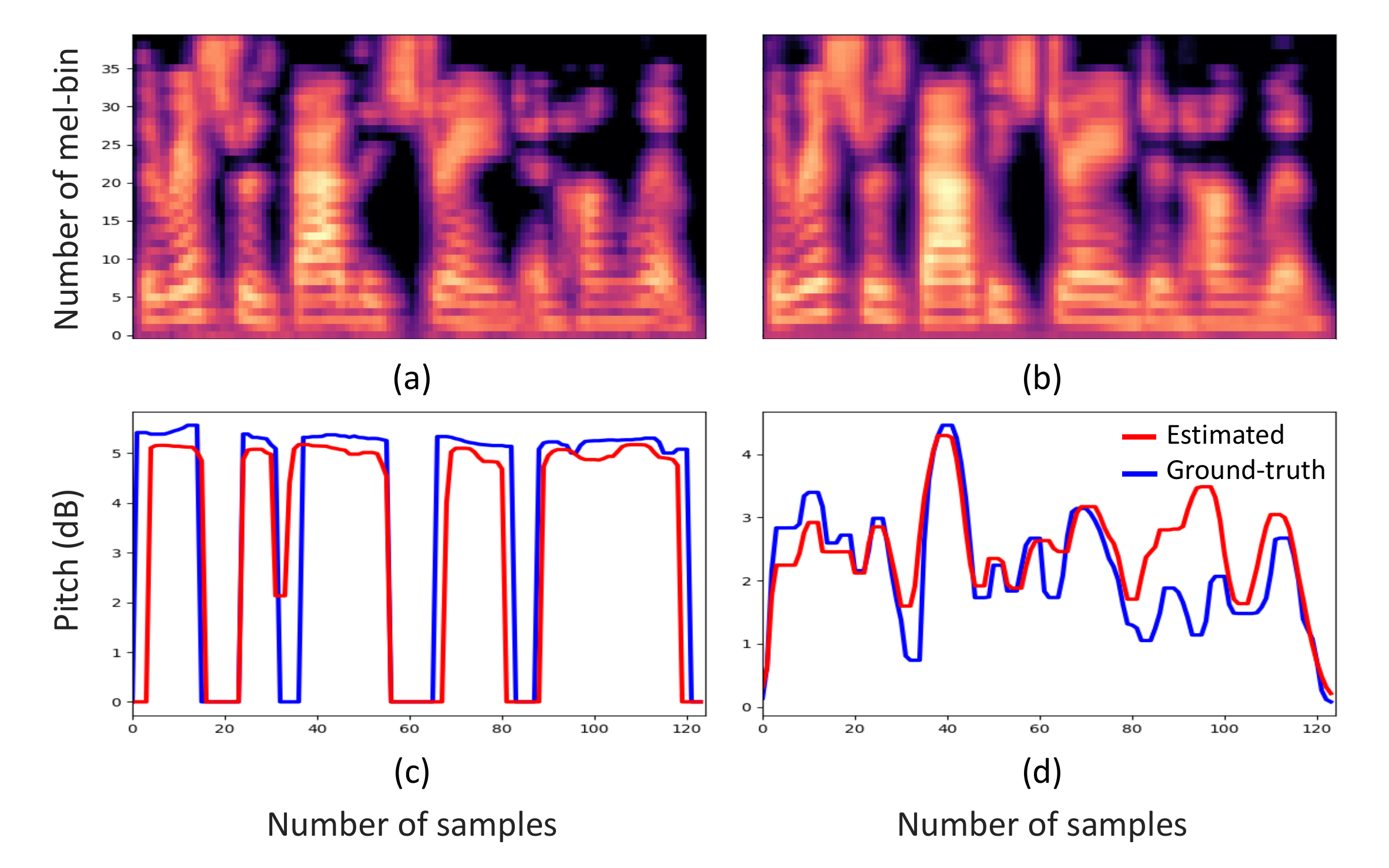}}
\caption{Mel spectrum, estimated pitch and energy trajectory for sentence \textit{"The wrist was badly strained and hung limp."}. (a): Target spectrum. (b): Synthetic spectrum from proposed model. (c) Pitch trajectory. (d) Energy trajectory. The blue line: target. Red line: estimated trajectory}  
\label{fig:pitch_energy}
\end{figure}

\subsection{Analysis of style embedding}
We plot ground-truth and synthesized mel-spectrum obtained from three systems in Figure~\ref{fig:analysis} to confirm the quality of generated mel-sectrum. 
Spectrum in (b) shows shorter length of formant compared with target spectrum. Spectrum in (c) shows blurry boundary between formant. Spectrum (d) displays visually the closest figure among three the ATS systems.
To investigate the effectiveness of style embedding more specifically, we visualize the estimated pitch and energy in Figure~\ref{fig:pitch_energy}. 
We can confirm that both the estimated pitch and energy follow up reference well. From the results, we prove that the style embedding exactly captures the speaker style information from the articulatory features.

\vspace{-8pt}
\section{Conclusion}
\vspace{-4pt}
\label{sec:conclusion}
In this paper, we introduced a novel high-quality speech synthesizer using articulation signal (EMA) in a multi-speaker scenario. 
By estimating the style embedding directly from EMA, which is definitely modeling pitch and spectrum energy components, we generated not only intelligible but also speaker style representative sound. 
Further fidelity of synthetic sound became increase by informing the speaker information to the network using a speaker embedding. 
As a model architecture, we adopted CNN and transformer attention layer based style encoder and mel-spectrum generator to synthesize high-quality speech from articulations. 
All experimental results proved that our proposed model achieves higher performance in terms of both of objective and subjective measurements.

\vspace{-12pt}
\section{Acknowledgements}
\label{sec:acknowledgements}

This work was supported by the project `Alchemist Brain to X (B2X) Project' through
the Ministry of Trade, Industry and Energy (MOTIE), South Korea, under Grant 20012355 and NTIS 1415181023. 

\vfill\pagebreak

\clearpage
\bibliographystyle{IEEEtran}
\small
\bibliography{refs}

\begin{thebibliography}{10}
\providecommand{\url}[1]{#1}
\csname url@samestyle\endcsname
\providecommand{\newblock}{\relax}
\providecommand{\bibinfo}[2]{#2}
\providecommand{\BIBentrySTDinterwordspacing}{\spaceskip=0pt\relax}
\providecommand{\BIBentryALTinterwordstretchfactor}{4}
\providecommand{\BIBentryALTinterwordspacing}{\spaceskip=\fontdimen2\font plus
\BIBentryALTinterwordstretchfactor\fontdimen3\font minus
  \fontdimen4\font\relax}
\providecommand{\BIBforeignlanguage}[2]{{%
\expandafter\ifx\csname l@#1\endcsname\relax
\typeout{** WARNING: IEEEtran.bst: No hyphenation pattern has been}%
\typeout{** loaded for the language `#1'. Using the pattern for}%
\typeout{** the default language instead.}%
\else
\language=\csname l@#1\endcsname
\fi
#2}}
\providecommand{\BIBdecl}{\relax}
\BIBdecl

\bibitem{chen2021ema2s}
Y.-W. Chen, K.-H. Hung, S.-Y. Chuang, J.~Sherman, W.-C. Huang, X.~Lu, and
  Y.~Tsao, ``Ema2s: An end-to-end multimodal articulatory-to-speech system,''
  in \emph{2021 IEEE International Symposium on Circuits and Systems
  (ISCAS)}.\hskip 1em plus 0.5em minus 0.4em\relax IEEE, 2021, pp. 1--5.

\bibitem{gates1982current}
G.~A. Gates, W.~Ryan, J.~Cooper~Jr, G.~F. Lawlis, E.~Cantu, E.~Lauder, R.~W.
  Welch, and E.~Hearne, ``Current status of laryngectomee rehabilitation: I.
  results of therapy,'' \emph{American journal of otolaryngology}, vol.~3,
  no.~1, pp. 1--7, 1982.

\bibitem{denby2010silent}
B.~Denby, T.~Schultz, K.~Honda, T.~Hueber, J.~M. Gilbert, and J.~S. Brumberg,
  ``Silent speech interfaces,'' \emph{Speech Communication}, vol.~52, no.~4,
  pp. 270--287, 2010.

\bibitem{anumanchipalli2019speech}
G.~K. Anumanchipalli, J.~Chartier, and E.~F. Chang, ``Speech synthesis from
  neural decoding of spoken sentences,'' \emph{Nature}, vol. 568, no. 7753, pp.
  493--498, 2019.

\bibitem{kauffman1976articulation}
S.~A. Kauffman, ``Articulation of parts explanation in biology and the rational
  search for them,'' in \emph{Topics in the Philosophy of Biology}.\hskip 1em
  plus 0.5em minus 0.4em\relax Springer, 1976, pp. 245--263.

\bibitem{volaitis1992phonetic}
L.~E. Volaitis and J.~L. Miller, ``Phonetic prototypes: Influence of place of
  articulation and speaking rate on the internal structure of voicing
  categories,'' \emph{The Journal of the Acoustical Society of America},
  vol.~92, no.~2, pp. 723--735, 1992.

\bibitem{toda2004mapping}
T.~Toda, A.~W. Black, and K.~Tokuda, ``Mapping from articulatory movements to
  vocal tract spectrum with gaussian mixture model for articulatory speech
  synthesis,'' in \emph{Fifth ISCA Workshop on Speech Synthesis}, 2004.

\bibitem{aryal2013articulatory}
S.~Aryal and R.~Gutierrez-Osuna, ``Articulatory inversion and synthesis:
  towards articulatory-based modification of speech,'' in \emph{2013 IEEE
  International Conference on Acoustics, Speech and Signal Processing}.\hskip
  1em plus 0.5em minus 0.4em\relax IEEE, 2013, pp. 7952--7956.

\bibitem{picart2011continuous}
B.~Picart, T.~Drugman, and T.~Dutoit, ``Continuous control of the degree of
  articulation in hmm-based speech synthesis,'' in \emph{Twelfth Annual
  Conference of the International Speech Communication Association}, 2011.

\bibitem{gaddy-klein-2021-improved}
\BIBentryALTinterwordspacing
D.~Gaddy and D.~Klein, ``An improved model for voicing silent speech,'' in
  \emph{Proceedings of the 59th Annual Meeting of the Association for
  Computational Linguistics and the 11th International Joint Conference on
  Natural Language Processing (Volume 2: Short Papers)}.\hskip 1em plus 0.5em
  minus 0.4em\relax Online: Association for Computational Linguistics, Aug.
  2021, pp. 175--181. [Online]. Available:
  \url{https://aclanthology.org/2021.acl-short.23}
\BIBentrySTDinterwordspacing

\bibitem{kong2020hifi}
J.~Kong, J.~Kim, and J.~Bae, ``Hifi-gan: Generative adversarial networks for
  efficient and high fidelity speech synthesis,'' \emph{Advances in Neural
  Information Processing Systems}, vol.~33, pp. 17\,022--17\,033, 2020.

\bibitem{cao2022speaker}
B.~Cao, A.~Wisler, and J.~Wang, ``Speaker adaptation on articulation and
  acoustics for articulation-to-speech synthesis,'' \emph{Sensors}, vol.~22,
  no.~16, p. 6056, 2022.

\bibitem{cao2018articulation}
B.~Cao, M.~J. Kim, J.~R. Wang, J.~P. van Santen, T.~Mau, and J.~Wang,
  ``Articulation-to-speech synthesis using articulatory flesh point sensors'
  orientation information.'' in \emph{INTERSPEECH}, 2018, pp. 3152--3156.

\bibitem{taguchi2018articulatory}
F.~Taguchi and T.~Kaburagi, ``Articulatory-to-speech conversion using
  bi-directional long short-term memory.'' in \emph{Interspeech}, 2018, pp.
  2499--2503.

\bibitem{csapo2020ultrasound}
T.~G. Csap{\'o}, C.~Zaink{\'o}, L.~T{\'o}th, G.~Gosztolya, and A.~Mark{\'o},
  ``Ultrasound-based articulatory-to-acoustic mapping with waveglow speech
  synthesis,'' \emph{arXiv preprint arXiv:2008.03152}, 2020.

\bibitem{katsurada20_interspeech}
K.~Katsurada and K.~Richmond, ``{Speaker-Independent Mel-Cepstrum Estimation
  from Articulator Movements Using D-Vector Input},'' in \emph{Proc.
  Interspeech 2020}, 2020, pp. 3176--3180.

\bibitem{ren2020fastspeech}
Y.~Ren, C.~Hu, X.~Tan, T.~Qin, S.~Zhao, Z.~Zhao, and T.-Y. Liu, ``Fastspeech 2:
  Fast and high-quality end-to-end text to speech,'' \emph{arXiv preprint
  arXiv:2006.04558}, 2020.

\bibitem{wang2018style}
Y.~Wang, D.~Stanton, Y.~Zhang, R.-S. Ryan, E.~Battenberg, J.~Shor, Y.~Xiao,
  Y.~Jia, F.~Ren, and R.~A. Saurous, ``Style tokens: Unsupervised style
  modeling, control and transfer in end-to-end speech synthesis,'' in
  \emph{International Conference on Machine Learning}.\hskip 1em plus 0.5em
  minus 0.4em\relax PMLR, 2018, pp. 5180--5189.

\bibitem{min2021meta}
D.~Min, D.~B. Lee, E.~Yang, and S.~J. Hwang, ``Meta-stylespeech: Multi-speaker
  adaptive text-to-speech generation,'' in \emph{International Conference on
  Machine Learning}.\hskip 1em plus 0.5em minus 0.4em\relax PMLR, 2021, pp.
  7748--7759.

\bibitem{gulati2020conformer}
A.~Gulati, J.~Qin, C.-C. Chiu, N.~Parmar, Y.~Zhang, J.~Yu, W.~Han, S.~Wang,
  Z.~Zhang, Y.~Wu \emph{et~al.}, ``Conformer: Convolution-augmented transformer
  for speech recognition,'' \emph{arXiv preprint arXiv:2005.08100}, 2020.

\bibitem{sainburg2020finding}
T.~Sainburg, M.~Thielk, and T.~Q. Gentner, ``Finding, visualizing, and
  quantifying latent structure across diverse animal vocal repertoires,''
  \emph{PLoS computational biology}, vol.~16, no.~10, p. e1008228, 2020.

\bibitem{ljspeech17}
K.~Ito and L.~Johnson, ``The lj speech dataset,''
  \url{https://keithito.com/LJ-Speech-Dataset/}, 2017.

\bibitem{baevski2020wav2vec}
A.~Baevski, Y.~Zhou, A.~Mohamed, and M.~Auli, ``wav2vec 2.0: A framework for
  self-supervised learning of speech representations,'' \emph{Advances in
  Neural Information Processing Systems}, vol.~33, pp. 12\,449--12\,460, 2020.

\bibitem{rani2012error}
N.~U. Rani and P.~Girija, ``Error analysis to improve the speech recognition
  accuracy on telugu language,'' \emph{Sadhana}, vol.~37, no.~6, pp. 747--761,
  2012.

\end{thebibliography}

\end{document}